# Self-induced topological transitions and edge states supported by nonlinear staggered potentials

Yakir Hadad[1], Alexander B. Khanikaev[2,3], and Andrea Alu[1,*]

[1]Department of Electrical and Computer Engineering, The University of Texas at Austin, 1616 Guadalupe St., UTA 7.215, Austin, TX 78701, USA

[2]Department of Physics, Queens College of The City University of New York, Queens, NY 11367, USA

[3]Department of Physics, The Graduate Center of The City University of New York, New York, NY 10016, USA

*The canonical Su-Schrieffer-Heeger (SSH) model is one of the basic geometries that have spurred significant interest in topologically nontrivial bandgap modes with robust properties. Here, we show that the inclusion of suitable third-order Kerr nonlinearities in SSH arrays opens rich new physics in topological insulators, including the possibility of supporting self-induced topological transitions based on the applied intensity. We highlight the emergence of a new class of topological solutions in nonlinear SSH arrays, localized at the array edges. As opposed to their linear counterparts, these nonlinear states decay to a plateau with non-zero amplitude inside the array, highlighting the local nature of topologically nontrivial bandgaps in nonlinear systems. We derive the conditions under which these unusual responses can be achieved, and their dynamics as a function of applied intensity. Our work paves the way to new directions in the physics of topologically non-trivial edge states with robust propagation properties based on nonlinear interactions in suitably designed periodic arrays.*

1. Introduction

Driven by recent advances in understanding topological phases in condensed matter physics [1]-[10] there has been a significant interest in engineering or emulating topological states in classical [11]-[17], and bosonic [18]-[22] systems. While other classes of topological order, e.g., with time reversal symmetry broken by an applied magnetic field, or symmetry-protected topological phases, have been successfully demonstrated in classical systems, nonetheless, due to their different statistics, bosonic systems lack one important class of topological protection stemming from time-reversal symmetry. This limitation has spawned activity to look into other possible mechanisms to induce topological phases, which provided new classes of topological order for bosons, and expanded the periodic table of topological classifications [23]-[24].

To date, most investigations in the area of classical topological insulators have been limited to linear phenomena. While some preliminary studies of nonlinear effects in topological systems have been recently presented [25]-[26], the question of whether nonlinearities may induce topological order remains unanswered. Here we investigate the most basic nonlinear topological classical system, based on the Su-Schrieffer-Heeger (SSH) model [27] loaded with a nonlinear staggered potential. We demonstrate that a suitable form of nonlinearity can result, under specific conditions, in self-induced transitions [28] from trivial to nontrivial topological states, accompanied by the emergence of self-trapping topological edge solitons, confined to the edges of the nonlinear SSH array.

Despite its simplicity, the SSH model has attracted extensive attention in the past, due to its rich phenomena, including topological excitation, fractional charge, and nontrivial edge states [29]-[33]. Recent developments in topological insulators have reignited interest in the SSH model as it represents the simplest example of a 1D topological insulator with particle-hole and chiral symmetries (BDI class) and it can host Majorana fermions [34]-[37]. In the context of classical waves, the SSH model can describe a variety of classical systems, including electromagnetic and acoustic 1D arrays of coupled resonators [38]-[40], indicating that these classical systems can host topological edge states. This opens new avenues to study topological order in classical systems in which the effective Hamiltonian can be precisely controlled, and, it becomes of special interest in the case studied here, when nonlinear interacting potentials are added to the picture. It is interesting that, while the SSH model has gained its original popularity in the context of nonlinear phenomena and solitons, recent developments in topological order brings back the relevance of the nonlinear SSH model, which turns out to be again instructive to understand the unique physics of nonlinear topological states, self-induced topological transitions and excitations described in the following.

## 2. Geometry of interest and analytical model

We investigate the nonlinear dynamics of a nonlinear SSH chain of dimers with intra-cell and inter-cell coupling coefficients, $v$ and $\kappa$, respectively, as shown in Fig. 1. The resonator amplitudes vector of the $n$-th dimer is given by $\Psi_n = [a_{1,n}, a_{2,n}]^T$, and the chain dynamics is described by the nonlinear Schrodinger equation

$$i\frac{d\Psi_n}{dt} = \mathbf{\Omega}\Psi_n + \mathbf{K}_m(n)\Psi_{n-1} + \mathbf{K}_p(n)\Psi_{n+1} \tag{1}$$

where the matrix $\mathbf{\Omega} = [\omega_0, v; v, \omega_0]$ encapsulates the intra-cell dynamics, and $\omega_0$ and $v$ are respectively the self-resonance frequency of an isolated resonator, and the intra-cell coupling coefficient. It is important to mention that Eq. (1) can model a broad range of classical problems in temporal coupled mode theory, for instance arrays of coupled optical and acoustic cavities, as well as circuit resonators [41]-[42]. The inter-cell coupling in Eq. (1) is described by the matrices $\mathbf{K}_m(n) = [0, \kappa_0 + \alpha(|a_{1,n}|^2 + |a_{2,n-1}|^2); 0, 0]$ and $\mathbf{K}_p(n) = [0, 0; \kappa_0 + \alpha(|a_{1,n+1}|^2 + |a_{2,n}|^2), 0]$, which contain a linear coupling term $\kappa_0 > 0$ and a Kerr-like nonlinear coefficient $\alpha \geq 0$. Evidently, the solutions of Eq.

(1) depend on the mode amplitudes. In the following, the coupling coefficients $v, \kappa_0$ as well as the independent time variable $t$ and all other frequency variables in the text are normalized with respect to the resonator self-resonance frequency $\omega_0$. The mode amplitudes $a_{1,n}, a_{2,n}$ are also normalized such that $|a_{1,n}|^2 + |a_{2,n}|^2$ is the $n$-th dimer stored energy. Thereby, the Kerr coefficient $\alpha$ has units of inverse energy.

We start our analysis by investigating an infinite SSH chain as in Fig. 1, assuming that the nonlinearity is weak enough so that the modal solution can be approximated using the Bloch-like wave function

$$\Psi_n = \Psi_0 e^{in\varphi - i\omega t}. \tag{2}$$

Once Eq. (2) is plugged into Eq. (1), we obtain a non-linear eigenvalue problem that reads

$$\mathbf{H}(\varphi; \Psi_0) \Psi_0 = \bar{\omega}\, \Psi_0, \tag{3}$$

with $\bar{\omega} = \omega - \omega_0$ and $\mathbf{H}(\varphi; \Psi_0) = h_x \sigma_x + h_y \sigma_y$, where $h_x = v - \kappa(a_{1,0}, a_{2,0}) \cos\varphi$, $h_y = \kappa(a_{1,0}, a_{2,0}) \sin\varphi$, $\kappa(a_{1,0}, a_{2,0}) = \kappa_0 + \alpha(|a_{1,0}|^2 + |a_{2,0}|^2)$, and $\sigma_{x/y}$ are the Pauli matrices. In contrast to the linear chain, the solution (2) is subject to a dispersion that depends also on the mode amplitudes $\Psi_0$.

### 3. Properties of the infinite nonlinear SSH array

By substituting Eq. (2) into Eq.(1), we can solve numerically the nonlinear chain dispersion (see Appendix A). Quite interestingly, at the edges of the first Brillouin zone, $\varphi = \pm\pi$, it is possible to derive an analytical closed-form solution for the bandgap frequency edges

$$\bar{\omega}_{bg} = \pm |v - \kappa_0 - \alpha I^2|, \tag{4}$$

where $I = \max_n \{\sqrt{|a_{1,n}|^2 + |a_{2,n}|^2}\}$ is the mode intensity, which, under the assumption (2), reduces to $I = \|\Psi_0\|$. The simple expression (4) provides relevant insights into the array bandgap, showing its evolution as a function of intensity. In particular, if $\nu > \kappa_0$, at low intensities the bandgap is open, as shown in Fig. 2(a), calculated for $\nu = 2.3 \times 10^{-3}$ and $\kappa_0 = 2.0 \times 10^{-3}$. After increasing the intensity, the gap will be closed at the threshold intensity $I_{th} = \sqrt{(\nu - \kappa_0)/\alpha} \approx 2.45$, and reopened again, as illustrated by Fig. 2(b).

Importantly, such bandgap evolution does not leave the system intact: as indicated in the figure it is easy to prove that the system experiences a unique, self-induced topological transition, switching from trivial to non-trivial topological properties, as the modal intensity is increased. In order to prove this claim one needs to calculate the topological invariant of the system, the winding number, as a function of the eigenvector intensity $I$. To this end, we define the state vector $\mathbf{h}(\varphi;\Psi_0) = h_x - ih_y$ that rotates and completes a full circle, centered at $(\nu, 0)$ in the complex $(h_x, h_y)$ plane as the phase $\varphi$ varies along the first Brillouin zone. The winding number $W$ is defined by [43]

$$W = \frac{1}{2\pi i} \int_{-\pi}^{\pi} d\varphi \frac{d}{d\varphi} \ln \mathbf{h}(\varphi;\Psi_0) \ . \tag{5}$$

It can readily be shown that $W = 1$ or $0$ depending on whether $\mathbf{h}(\varphi;\Psi_0)$ encircles or does not encircle the origin of the complex plane. By increasing the intensity, the length of the vector $\mathbf{h}$ increases, and at a certain critical value starts encircling the origin, as illustrated in Fig. 2(c). For the given set of parameters, and with nonlinear Kerr coefficient $\alpha = 5 \times 10^{-5}$, the critical threshold intensity value at which the winding number experiences a jump from 0 to 1 and the systems changes its topological state is $I_{th} = 2.45$, as shown in Fig. 2 (d).

The description put forward in this section, while instructive, has the limitation of using global properties to describe a nonlinear response that is local by nature, since it is a function of the local intensity along the array. Particularly, in semi-infinite or finite structures the description based on a Bloch-like state (2) is not valid, as the amplitude along the chain cannot be uniform due to the presence of boundaries, yielding inhomogeneous non-linear coupling coefficients along the chain that break the periodicity. Yet, as seen in the following, this basic model powerfully captures important insights to explain the mechanisms observed in finite and semi-infinite nonlinear SSH arrays based on the geometry of Fig. 1.

### 4. Finite systems: the role of edges and boundaries

Studying truncated arrays is crucial to get more insights into the unusual topological response outlined in Fig. 2, particularly regarding the transition from trivial to non-trivial topology, and the possibility of emergence of edge states, which, from the bulk-boundary correspondence principle [44], are known to exist in linear systems with topological order characterized by $W = 1$. To this end, we consider a nonlinear SSH chain of $N = 40$ dimers, under a harmonic excitation with time dependence $e^{-i\omega t}$. Using the numerical procedure discussed in Appendix B, the resultant $2N \times 2N$ nonlinear eigenvalue problem was solved iteratively for the lowest-order eigenvalues, and we show the results in Fig. 3. We use the same coefficients $v, \kappa_0, \alpha$ used for the calculation in Fig. 2 relative to the infinite array. In Fig. 3(a), a set of eigenfrequencies of the linear (low-intensity) problem is shown, and contrasted with the dispersion of the infinite structure. All eigenfrequencies, as expected, are located outside the bandgap in the passband. In Fig. 3(b), we then explore the evolution of the two smallest eigenfrequencies (marked by bold red + in Fig. 3(a)), as we increase the eigenvector intensity, and the position of the bandgap edges, consistent with Fig. 2b, is plotted as brown dashed lines. At low intensities, before the bandgap closes, the trajectories of the lowest eigenfrequencies follow the bandgap from outside. Remarkably, exactly at the critical threshold value of intensity required to close the bandgap $I_{th} = 2.45$, an abrupt change in eigenfrequency trajectory is observed (refer to the inset in Fig. 3(b) for a zoom in this critical region), indicating the onset of a different response. Starting from this intensity on, the eigenfrequency trajectories enter the bandgap, and gradually converge to $\bar{\omega} = 0$, as one would expect for an edge state.

In order to get further understanding into this transition, we explore how the corresponding eigenvectors evolve as the intensity increases. In Fig. 3(c), we show the absolute value of the eigenvector distribution versus the dimer index $n$ and the modal intensity $I$. At very low intensities, a typical standing-wave is formed across the finite array, a cosine-like bulk solution. Its profile can be also seen in Fig. 3(d) for the case $I \simeq 0$. As the intensity is increased, however, the mode profile gradually deforms, and it acquires a hyperbolic-cosine-like shape that eventually forms strongly decaying tails located at the two opposite ends of the chain. The mode profile in this case can be better seen in Fig. 3(e) for a high normalized intensity $I \simeq 60$. The figure allows us to observe the gradual transition from trivial bulk modes near the bandgap to localized edge modes in the non-trivial bandgap, as the intensity grows. Interestingly, and very different from edge modes in linear topological insulators, the localized edge modes observed in Fig. 3(e) do not appear to decay to zero in the bulk, but they converge to a constant level of intensity, a feature that will be explained in the following. Nevertheless, as seen in Fig. 3(b), the eigenfrequencies converge simultaneously to $\bar{\omega} = 0$, consistent with the fact that, as the intensity increases, the finite chain solution can be described as a combination of two nearly uncoupled solutions of the corresponding semi-infinite problem, one for each edge, with resonances very close to the individual resonator case.

To gain further insights into this unusual edge response, we explore the analytical solution of the semi-infinite array. This solution is derived in detail in Appendix C, where we derive the analytical condition describing an edge-state decaying from the left edge ($n = 1$), which is shown to require $\bar{\omega} = 0$, $a_{2,n} \equiv 0, \forall n$. Similarly, the opposite edge-state requires $a_{1,n} \equiv 0$. The eigenvector magnitude of the semi-infinite solution is shown by the red-dashed curve in Fig. 3(e), demonstrating a very good agreement with the numerical solution of the finite chain. Using this solution, we are also able to analytically predict the plateau level to which the solution converges as it decays in the bulk of the array, which depends on the array properties through the simple relation

$$|a_{1,n}| = \sqrt{\frac{\nu - \kappa_0}{\alpha}} \tag{6}$$

This result makes perfect physical sense since this amplitude corresponds to the local threshold intensity that supports a zero bandgap width, as was shown after Eq. (4). The same is true for the edge state localized on the right edge with $|a_{2,n}| = \sqrt{(\nu - \kappa_0)/\alpha}$ and $a_{1,n} = 0$. Zero bandgap implies zero attenuation, thereby keeping the intensity unchanged, and yielding a self-sustained balance with constant amplitude that avoids the creation of a topological domain wall within the array. This observation demonstrates the importance of the local nature of the nonlinear problem at hand.

Figure 4 shows the effect of increasing the finite chain length to outline the transition from a semi-infinite array to a finite array supporting localized edge modes. In panel (a), the evolution of the lowest-order eigenfrequency as a function of intensity is shown for different chain lengths. The drastic change in dynamics at the threshold intensity $I_{th} \approx 2.45$ is evident. As $N$ increases, the intensity dependence converges to the semi-infinite case, especially in the case of high intensities deep in the edge-state regime, as shown in (b). This is an important feature since, as opposed to linear topological edge states, which are exponentially decreasing to zero, the edge states that we find decay to a non-zero plateau along the array. It is remarkable that, despite this finite coupling, the two edge states are orthogonal, as discussed in Appendix C. It should be also pointed out that the edge state of the semi-infinite chain has frequency $\bar{\omega} = 0$. The frequency splitting observed in Fig. 4(a) and 4(b) is a consequence of the unavoidable coupling between the finite chain edges. In the linear problem, this coupling reduces as the mode is more confined, and the splitting shrinks as the chain length increases. However, in the nonlinear model studied here, there is an additional important factor that controls the coupling, which is the intensity dependence. Increase in intensity leads to increased mode attenuation, and thereby reduced coupling, in such a way to compensate for a decrease in the chain length. This effect is demonstrated in Fig. 4(c), where for each chain length we plot the corresponding edge state intensity that is required to maintain a constant frequency splitting $\bar{\omega} = 2 \times 10^{-6}$, and thereby constant coupling.

## 5. Nonlinear temporal dynamics

The discussion so far has been based on a frequency domain analysis of the nonlinear SSH array, consistent with a traditional condensed matter physics approach to the problem. However, it is evident that the nonlinear topological states presented in this work have unique dynamical properties that depend on how a certain level of intensities is reached in the array. It may be possible that some of the states considered here may not even be reached with a physical excitation. In this section, therefore, we study the excitation problem of the system to provide further insights into the existence and dynamics of self-induced topological transitions and topologically non-trivial edge states with robust properties. In this case, we use a time-domain analysis to examine the entire temporal spectra, providing also insights into the evolution of the array response upon realistic excitation schemes. Thereby, we solve a time-domain excitation problem described by the non-homogenous equation obtained when a source $\mathbf{S}_n(t)$ coupled through a linear coupling coefficient $\xi$ (normalized to $\omega_0$) is added to the right-hand-side of Eq.(1). The source is given by a discretized Gaussian profile with temporal width $\tau$

$$\mathbf{S}_n = \delta_{n-1}[S_{in}(t),0]^T, \quad S_{in}(t) = S_0 e^{-i\omega t} e^{-(t-t_0)^2/\tau^2} \tag{7}$$

In the equation, $\delta_{n-1}$ stands for the Kronecker delta, thus indicating that the source is coupled only to the element $a_{1,1}$ of the first dimer, and $S_0^2$ has energy units as $|a_{1,n}|^2, |a_{2,n}|^2$. In addition to a Gaussian profile, the source (7) is modulated in frequency, and its parameters are chosen such that the generated

Fourier spectrum covers the entire spectral window of interest, i.e., the entire passband of the chain. The dynamic solution for this form of excitation is, therefore, equivalent to the system's response to an incident broadband impulse. The left (right) column in Fig. 5 shows results for weak $\xi S_0 = 5 \times 10^{-3}$ (strong, $\xi S_0 = 10^{-1}$) excitation amplitude. In Fig. 5(a) the time domain response of dimers at the left edge, center, and right edge are shown. The amplitudes in the different dimers are roughly equal, and the time domain signals exhibit an echo-like shape indicating the presence of multiple reflections at the two edges, as expected in the excitation of a standing-wave bulk mode.

Markedly different is the response for a strong input excitation, shown in Fig. 5 (b) ($\xi S_0 = 10^{-1}$). It is readily seen that the first dimer is excited much more strongly than all dimers in the bulk and, in addition, no signal bounces back from the edge truncations, indicating the absence of multiple reflections. This agrees well with the expected localized behavior of an edge state. The corresponding spectral content, as given by the power spectrum of $a_{1,1}(t)$, is shown in Fig. 5 (c) and (d) for the two cases, respectively. In (c) a clear picture of a passband, stopband, and multiple sharp peaks that correspond to various discrete states supported by the finite linear chain are found. However, in (d) we observe a single dominant peak at the frequency $\bar{\omega} = 0$, located right at the center of the bandgap. This is another clear manifestation of a self-trapping of the edge state in the nonlinear SSH chain, fully consistent with our eigen-mode analysis above.

Besides the temporal and spectral contents, the field distribution is another important indication of the self-trapping of the edge state due to the nonlinear topological transition described in this paper. This is shown in Fig. 5 (e) and (f), in which we plot the peak amplitude values at each dimer. In (e) we clearly see the presence of bulk modes, whereas in (f) an exponentially decaying edge state is demonstrated, with a plateau similar to the one predicted by our frequency domain eigenvalue analysis in Fig. 3(e). Finally, to fully reveal the topological nature of the self-trapped edge states, it is instructive to explore the topology of the chain by analyzing the topological invariant $W$. Given the nonlinearity, the array cannot be considered any longer periodic in the usual, linear sense. Therefore, in Fig. 5 (g) and (f), we plot the "local winding number" $W_n$ at each node of the array, which represents the winding number corresponding to an infinite chain with coupling coefficients as that of the $n$ dimer for the given intensity. In the weak excitation case, this quantity is identically zero all along the array (Fig. 5(g)), indicating that the entire structure is topologically trivial, whereas in the strong excitation the situation is opposite (Fig. 5(f)), and the plateau intensity all along the array bulk ensures that the winding number stays at unity.

It is interesting to investigate more carefully the dynamics of the topological transition between trivial and non-trivial states in the system. Figure 6(a) shows the local winding number versus dimer index and the input signal intensity. One can see that at low input intensities the entire chain is topologically trivial. As the input intensity reaches a threshold $\xi S_0 \approx 0.02$, parts of the chain in the proximity to the excitation point experience a transition from topologically trivial to non-trivial regime, leading to the emergence of a localized resonance at $\bar{\omega}=0$. This transition then builds up very abruptly as an avalanche effect, due to a focusing effect around the array edges, which ends abruptly when the input intensity is slightly increased above the threshold value, and the entire structure becomes topologically non-trivial. As an illustration, in Fig. 6 (b) we plot the frequency at which the strongest excitation takes place. Interestingly, once a small part of the chain becomes topologically non-trivial ( $\xi S_0 \approx 0.02$ ), the strongest excitation frequency jumps from outside the bandgap right into its center $\bar{\omega}=0$, and an edge state is excited, in agreement with the previous description. In Fig. 6(c), we show the peak amplitude at the frequency of strongest excitation normalized to the input signal. As the input intensity increases, the power that is delivered to the edge-state increases as well, yielding a strong power concentration in the edge-state inside the bandgap, which can become orders of magnitude larger than the input excitation.

## 6. Conclusions

In this work, we introduced and discussed the rich platform offered by nonlinear asymmetric dimer (SSH) arrays in the context of topological properties. We demonstrated that such nonlinear SSH arrays exhibit self-induced nonlinear topological phase transitions, which are accompanied by an emergence of self-trapping edge states, supported by an effective structural modification of the topological properties of the array as a function of the amplitude. We explored this new class of soliton-like topological edge modes in semi-infinite and finite SSH chains, both in frequency and in time domain, and revealed the nature and dynamics of their self-trapping dynamics. This new class of nonlinear topological responses differs from conventional edge states in linear systems, as they locally depend on the effective environment induced by the non-linearity. The very emergence of such topological transitions with edge state demonstrates that nonlinear effects can serve as a new important class of mechanisms of topological order in classical systems, which can be readily extended to 2D and 3D arrays.

**Appendix A**

*Dispersion calculation of the infinite non-linear SSH chain:* For given modal intensity, the non-linear eigenvalue problem in Eq. (3) should, in general, be solved iteratively. However, in light of the simple non-linear model we assume, the iterative process can be replaced by direct calculation. Recall that in the case of weak nonlinearity we assume Bloch-like solution and define the intensity by $I = \|\Psi_0\|$. Assuming that $I$ is known in Eq. (3), the corresponding eigenfrequencies can be calculated through,

$$\bar{\omega} = \pm\sqrt{v^2 + 2v\cos\varphi\left[\kappa_0 + \alpha I^2\right] + \left[\kappa_0 + \alpha I^2\right]^2}. \tag{8}$$

The dispersion $\bar{\omega}(\varphi)$, $\varphi \in [-\pi, \pi]$, depends on the intensity. Once the frequency-phase relation is known for given intensity, the value of each amplitude can be also calculated using

$$a_{1,0} = -\bar{\omega}^{-1}\left\{v + \left[\kappa_0 + \alpha I^2\right]e^{-i\varphi}\right\}a_{2,0} \tag{9}$$

*Formula for the bandgap:* The bandgap takes place at $\varphi = \pm\pi$, therefore Eq. (8) yields

$$\bar{\omega}_{bg} = \pm\left|v - \kappa_0 - \alpha\left(|a_{1,0}|^2 + |a_{2,0}|^2\right)\right| \tag{10}$$

which can also be written in the form of Eq. (4). Note that, if we assume that $a_{2,0}$ is real, then by Eq. (9) $a_{1,0}$ will be necessarily real as well, and therefore in this case the absolute value sign on the amplitudes in Eq. (10) is not required. Lastly, note that if we use Eq. (10) in Eq. (9) with $\varphi = \pm\pi$ we get $a_{1,0} = \pm a_{2,0}$ which is true only in the case of propagating solutions, very close to the bandgap.

**Appendix B**

*Solution of the nonlinear eigenvalue problem of a finite chain:* Assume that we have $N$ dimers in a finite chain. Then, Eq. (1) reduces to a nonlinear $2N \times 2N$ matrix eigenvalue problem of the form

$$\mathbf{H}(\Psi)\Psi = \omega\Psi \qquad (11)$$

with $\Psi = [\Psi_1; \Psi_2; ..; \Psi_N]$ a $2N \times 1$ column vector, and where $\Psi_n = [a_{1,n}, a_{2,n}]^T$ as defined in the main text. When solving Eq.(11), we simultaneously look for an eigenfrequency $\omega$ and eigenvector $\Psi$ that satisfy this equation. In the linear case $\mathbf{H} \neq \mathbf{H}(\Psi)$ and the solution is straightforward as the spectrum can be determined from the secular equation $\det(\mathbf{H}) = 0$. However, in our case the problem is nonlinear and the Hamiltonian depends on the eigenvector. Therefore, the problem cannot in general be solved in any other way but iteratively. For this iterative procedure, an initial guess is required. Two possible candidates are the eigenvector of the weakly nonlinear problem, namely, with very low intensity, or the high-intensity solution that can be approximated by the analytic solution for the semi-infinite chain and given in Appendix C. The letter approach is the preferred way, as this solution is very distinct in its shape as compared with the former solution of the problem, and it is indeed closer to the true solution of the finite-chain problem. Finally, note that there is an important difference between the nonlinear eigenvalue problem we deal with and a linear eigenvalue problem. In the latter, an eigenvector that is associated with an eigenvalue is defined up to a scaling constant. In our case, however, an eigenvector that corresponds to a particular eigenfrequency is uniquely defined. And if scaled by a constant factor it does not anymore represent an eigenvector of Eq. (11).

**Appendix C**

*Calculating the mode profile of an edge-state:* Based on our numerical solutions both in the frequency and in the time domain, we expect to find edge state solutions at $\bar{\omega} = 0$. This is confirmed by the convergence of the lowest eigen-frequencies of the finite problem to the $\bar{\omega} = 0$ axis. This implies that we should expect to have solutions to the semi-infinite problem that can exhibit a vanishing coupling in the finite case. This can be achieved if, for the solution decaying from the left edge towards the right, one has $a_{2,n} = 0$, whereas for the solution localized to the opposite edge one should have $a_{1,0} = 0$. We focus on the former, noting that the latter can be readily obtained by the simple mirror reflection operation. We assume real solution, therefore $|a_{1,n}|^2 = a_{1,n}^2$, in which case the frequency-domain counterpart of Eq. (1) can be obtained by assuming the time-harmonic solution $\Psi_n(t) = \Psi_n(\omega)e^{-i\omega t}$ and Eq. (1) reduces to the cubic equation

$$a_{1,n+1}^3 + \frac{\kappa_0}{\alpha} a_{1,n+1} + \frac{v}{\alpha} a_{1,n} = 0 \tag{12}$$

This equation should be solved for $a_{1,n+1}(a_{1,n})$ in order to get a recursion type solution of the non-linearly induced edge state. To this end, we start by defining simpler equation parameters $p = -\kappa_0/\alpha$ and $q = -a_{1,n} v/\alpha$, followed by a change of variable

$$a_{1,n+1} = \sqrt[3]{w} + \frac{p}{3\sqrt[3]{w}} \tag{13}$$

which turns Eq. (12) into a much simpler quadratic equation

$$w^2 - qw + \frac{p^3}{27} = 0 \tag{14}$$

Eq. (14) has two real roots, one positive and one negative,

$$w = -\frac{v}{2\alpha} a_{1,n} \pm \frac{1}{2}\sqrt{\frac{v^2}{\alpha^2} a_{1,n}^2 + \frac{4}{27} \frac{\kappa_0^3}{\alpha^3}} \tag{15}$$

Any of these roots, when substituted back into Eq. (13), gives exactly the same three solutions of the cubic equation. One is real and the other two are complex. We are interested in the real solution only

(complex $a_{1,n}$ would contradict the assumption under which Eq. (12) was derived), therefore, we take the real root of $\sqrt[3]{w}$ in Eq. (13).

Using this exact edge state solution one can show that the (localized) edge state exists, namely that $|a_{1,n+1}| < |a_{1,n}|$ for all $n$, thus indicating that the amplitude attenuates away from the edge, only if

$$|a_{1,0}| > a_{1,\text{TH}} = \sqrt{\frac{\nu - \kappa_0}{\alpha}} \tag{16}$$

This result makes a lot of sense, since this is exactly the condition at which the nonlinear coupling coefficient $\kappa$ equals a value that exactly closes the bandgap according to Eq. (4), assuming that $a_{2,n} = 0$. Above this threshold the chain will be topologically non-trivial, enabling the excitation of an edge-state at $\bar{\omega} = 0$. Moreover, once the edge-state decays, the "local band gap" is gradually closed and eventually the edge state will stop decaying. This will happen, again, at the critical threshold amplitude in Eq. (16). Therefore, we conclude that the minimum edge-state amplitude possible in non-linearity induced topological insulators depends directly on the band-gap of the linear $|\nu - \kappa_0|$ problem and is inversely proportional to the non-linear Kerr coefficient $\alpha$. It is important to note that the mathematical solution given by Eq. (13) along with Eq. (15), represents a physical solution only when Eq. (16) is satisfied. In this case the solution is monotonically decreasing, whereas, if the initial intensity is lower than that the threshold value, the solution will be monotonically increasing, which is an unphysical behavior for a passive system. This proves that the edge state exists, and can be excited, only for sufficiently high excitation intensities, as it is demonstrated in the context of time-domain solutions presented in Fig. 4 and Fig. 5 and discussed in the main text.

**Acknowledgments**

This work was supported by the National Science Foundation, and the Air Force Office of Scientific Research.

*To whom correspondence should be addressed: alu@mail.utexas.edu

**Figures**

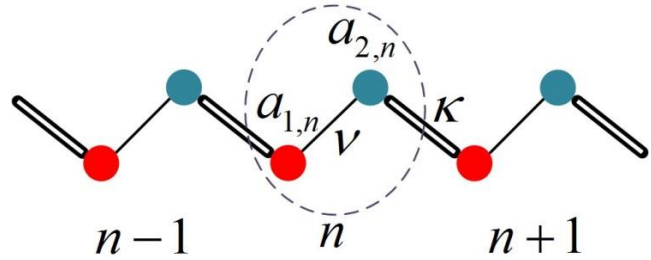

**Figure 1. Geometry of interest: nonlinear SSH model.** Each unit cell consists of two resonators, red and blue, with same resonance frequency $\omega = \omega_0$ ($\bar{\omega} = 0$). The intra-cell coupling coefficient is $v$, kept constant, whereas the inter-cell coupling coefficient $\kappa$ depends on the intensities in the two resonators connected by the bond.

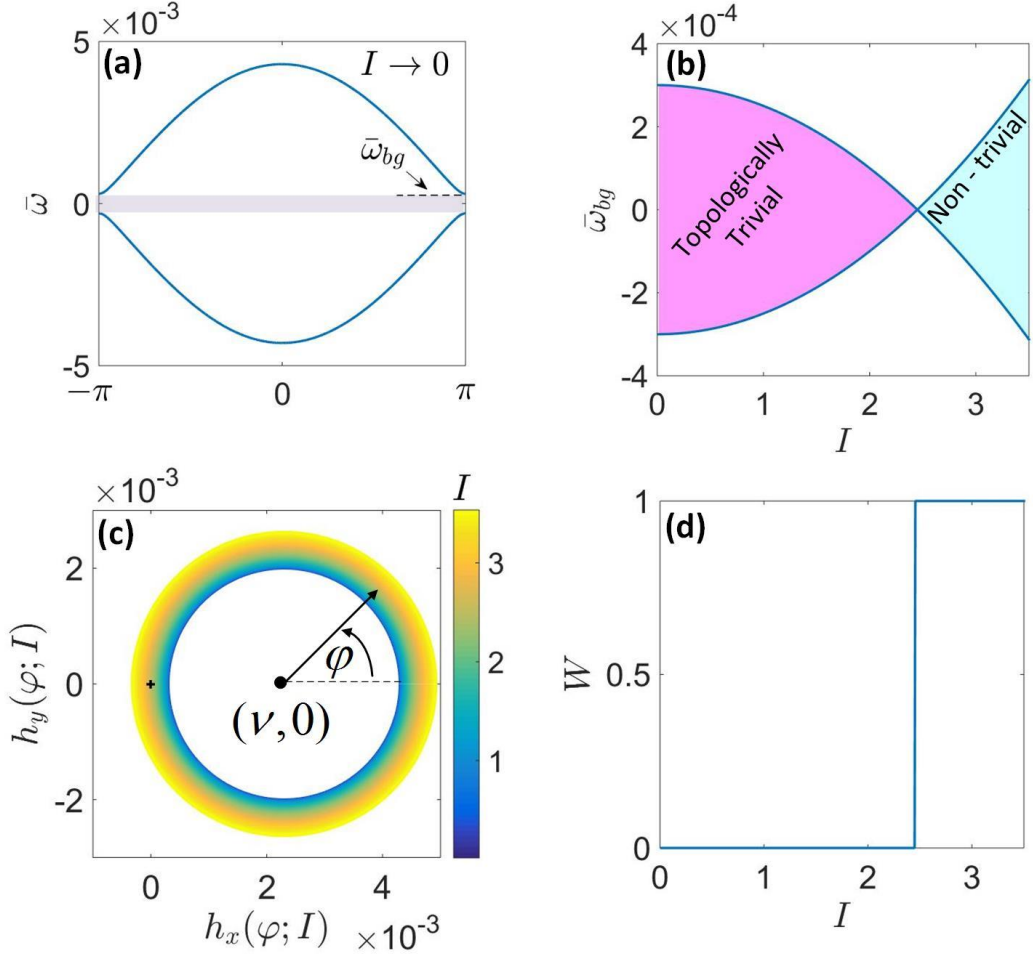

**Figure 2. Infinite nonlinear SSH chain.** (a) For negligible intensity $I \to 0$, the system reduces to the linear problem. Here $v = 2 \times 10^{-3}$ and $\kappa_0 = 2.3 \times 10^{-3}$, such that the dispersion has a topologically trivial bandgap. (b) The bandgap width is tuned by intensity. With non-linear Kerr coefficient $\alpha = 5 \times 10^{-5}$ the bandgap is closed and reopened at modal intensity $I_{th} \approx 2.45$. (c) The edge of the eigenvector of the operator $\mathbf{H}(\varphi; \Psi_0)$ is plotted for different intensities; each is represented by a different color shown in the color bar. The topological number, or winding number $W = 0$ when the vector does not encircle the origin (indicated by the + sign) and $W = 1$ otherwise. An intensity of $I_{th} \approx 2.45$ is the threshold value at which the circular trajectory created by the edge of the eigenvector encircles the origin. (d) $W$ plotted vs intensity $I$, illustrating the topological transition at $I_{th} \approx 2.45$ for which the bandgap closes and reopens.

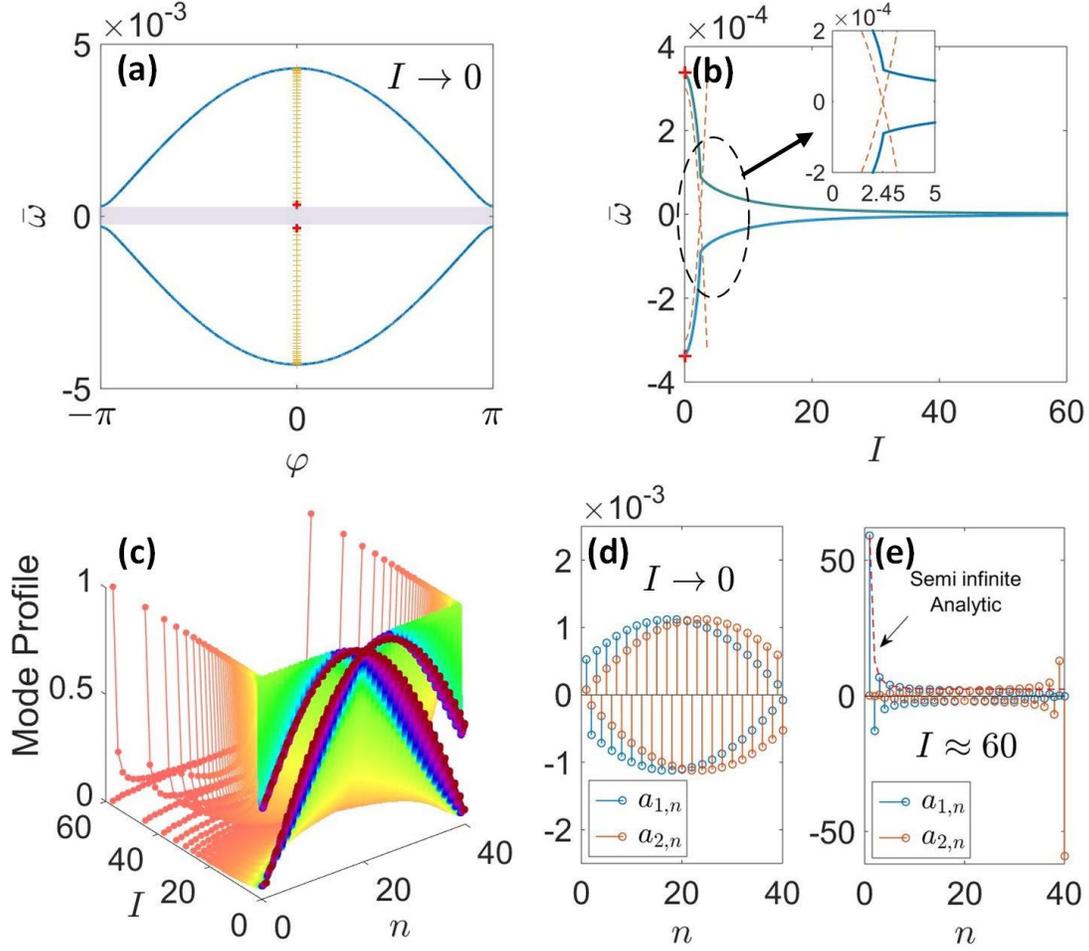

**Figure 3. Frequency domain analysis of a finite nonlinear SSH chain.** (a) Topologically trivial linear chain of $N = 40$ dimers, with 40 pairs of eigenfrequencies; all eigenfrequencies are located inside the passband of the infinite chain. (b) The trajectory of the smallest eigenfrequencies (marked by bold red + sign in (a)) versus intensity. The bandgap versus intensity is plotted in dashed lines. At the threshold intensity, for which the bandgap closes and reopens, an abrupt change in eigenfrequency trajectory is clearly observed and the eigenfrequencies steadily converge to $\bar{\omega} = 0$. (c) The evolution of the corresponding mode profile versus intensity demonstrating a transition from standing wave to exponentially decaying solutions localized at the edges. (d) Snapshot of the mode profile at zero intensity. (e) Snapshot of the localized edge state at the highest mode intensity. The analytical solution for the semi-infinite nonlinear SSH chain is plotted as a red dashed curve.

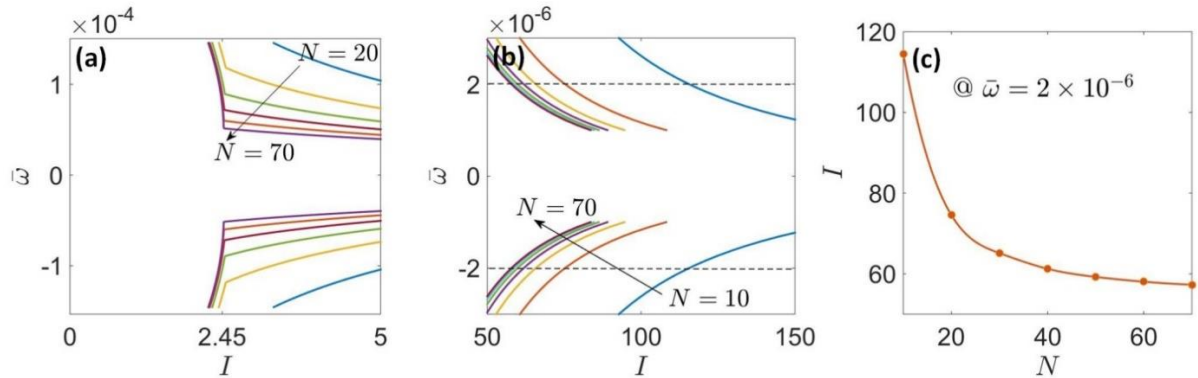

**Figure 4. Finite chain response versus array length. (a)** Evolution of the lowest order eigenfrequency versus intensity for different chain lengths. The threshold for which the dynamics of the eigenfrequency evolution changes, from bulk state to edge state, depends on the infinite chain bandgap crossing point. **(b)** The same as (a) but for high intensities, deep in the edge state regime. Like in (a), as the chain length increases the eigenfrequency trajectory tends to converge, as expected in an edge state response. **(c)** Maximal intensity versus chain length at a fixed eigenfrequency $\bar{\omega} = 2 \times 10^{-6}$. In the finite chain, the eigenfrequencies converge to $\bar{\omega} = 0$, but never reach that limit due to the unavoidable coupling between the two edges of the finite chain. If the coupling coefficient is kept constant, i.e., the frequency splitting $\bar{\omega}$ is constant, while the chain length is reduced, the modal intensity increases in order to decrease the attenuation rate.

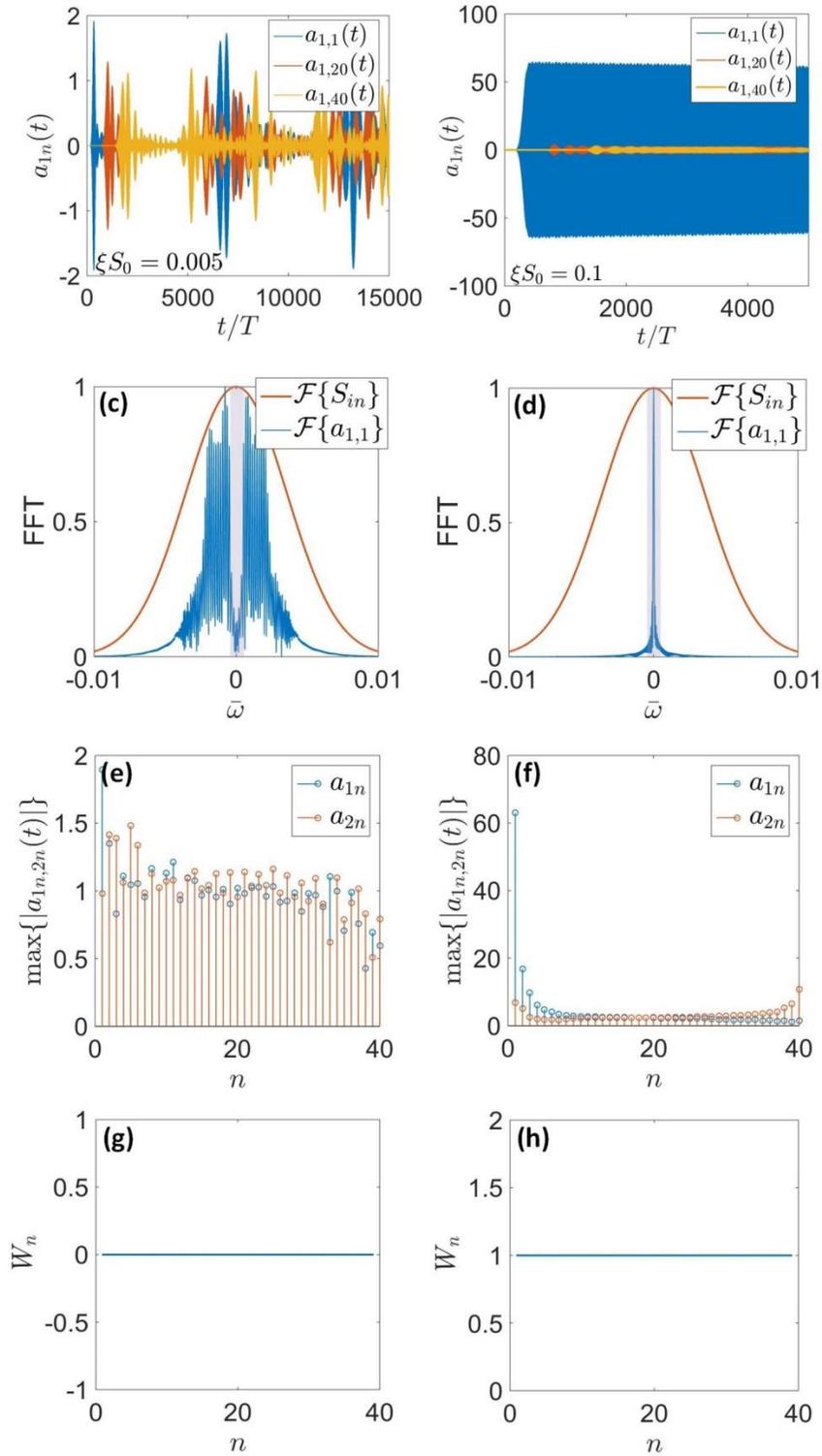

**Figure 5. Time-domain analysis of a finite** $N = 40$ **chain.** Left column: excitation with a modulated Gaussian pulse with amplitude $\xi S_0 = 5 \times 10^{-3}$; Right column: same but with higher excitation amplitude

$\xi S_0 = 10^{-1}$. **(a)** Time-domain response of the first resonator in three dimers, located at the left edge, center, and right edge of the chain. The amplitudes are nearly the same, and multiple reflections of the bulk mode are observed. **(b)** The same as (a) but for $\xi S_0 = 10^{-1}$. Excitation of the dimers in the center and at the right edge of the chain is negligible. No reflections are found, indicating the emergence of edge-states. **(c)** The Fourier (power) spectrum of the signals in (a). At low input intensity, the chain response is nearly linear, exhibiting a clear passband and a bandgap in the center. **(d)** Same as (c), but for $\xi S_0 = 10^{-1}$. A dominant excitation of an edge state in the center of the bandgap is evident. **(e)** Amplitude distribution along the chain, indicating excitation of a bulk mode. **(f)** Same as (e), indicating the excitation of an edge-state. **(g)** Local topological number calculated from the amplitudes in (e), indicating uniform topologically trivial chain. **(h)** Same as (g), indicating uniformly topologically non-trivial chain.

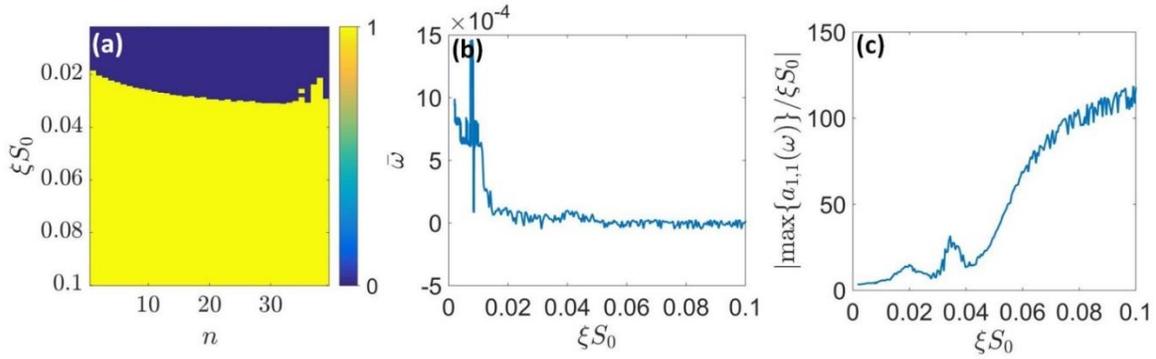

**Figure 6. Evolution of an edge-state through time-domain study.** (**a**) Local winding number versus input intensity and dimer index. (**b**) Frequency of the strongest excitation per given intensity. A sharp transition is observed at about $\xi S_0 = 0.02$, which is in agreement with the transition seen in the winding number. The spectrum peak is located on the $\bar{\omega} = 0$ axis if an edge-state is excited. (**c**) Spectrum density at peak amplitude normalized to the input signal amplitude. The graph indicates increasing power concentration in the edge state as the input intensity is increased.